\documentclass[twocolumn,aps,showpacs,prl,amsmath,amssymb,floatfix]{revtex4}
\usepackage{color}
\usepackage{graphicx}
\usepackage{dcolumn}
\usepackage{bm}
\usepackage{array}
\usepackage{float}
\usepackage{supertabular}
\usepackage{longtable}
\usepackage{mathrsfs}
\usepackage{wasysym}
\usepackage{amssymb}
\begin{document}

\title{Frustrated colloidal ordering and fully packed loops in arrays of optical traps}

\author{Gia-Wei Chern, C. Reichhardt, and C.~J. Olson Reichhardt}
\affiliation{Center for Nonlinear Studies and Theoretical Division, Los Alamos National Laboratory, Los Alamos, NM 87545, USA}  

\date{\today}

\begin{abstract}
We propose that a system of colloidal particles 
interacting with a honeycomb array of optical traps 
that each contain three wells
can be used to realize a fully packed loop model.
One of the phases in this system can be mapped to 
Baxter's three-coloring problem, 
offering an easily accessible physical realization of this problem.
As a function of temperature and interaction strength, 
we find a series of phases, including
long range ordered loop or stripe states, 
stripes with sliding symmetries, random packed 
loop states, and disordered states in which the
loops break apart. 
Our geometry could be constructed using ion trap arrays, 
BEC vortices in optical traps, or magnetic vortices
in nanostructured superconductors.     
\end{abstract}

 
\maketitle

{\em Introduction.}
There has recently been tremendous growth in the area 
of creating idealized systems
in which certain types of statistical mechanics models 
with and without geometric frustration can be physically realized, 
such as in nanomagnets~\cite{5,6,7}
and soft matter systems 
\cite{Brunner02,babic05,libal06,L,1,2,3,9,10,11,olson12}.
The key advantage of these systems is that they allow 
direct experimental access to the microscopic degrees of freedom.
One of the most active of these areas has been 
artificial spin ices created using
nanomagnetic arrays with square \cite{5} 
or hexagonal ordering \cite{6,7}, 
where ordered or frustrated states 
can occur that mimic real spin ice systems \cite{8}. 
Here, various types of excitations such as monopoles can arise, 
and the dynamics can be studied under an external field \cite{7}. 
There are many other statistical mechanics models 
that exhibit frustration effects,  including loop models 
such as the famous Baxter's three-coloring model~\cite{baxter70},
where only very limited work has been performed on proposed physical
realizations, all of which involve atomic degrees of freedom
\cite{moore04,castelnovo04,xu,you12}.
The nanomagnetic systems have certain constraints that
make it very difficult to realize many other types of statistical 
mechanics models of interest.

\begin{figure}
\includegraphics[width =3.5in]{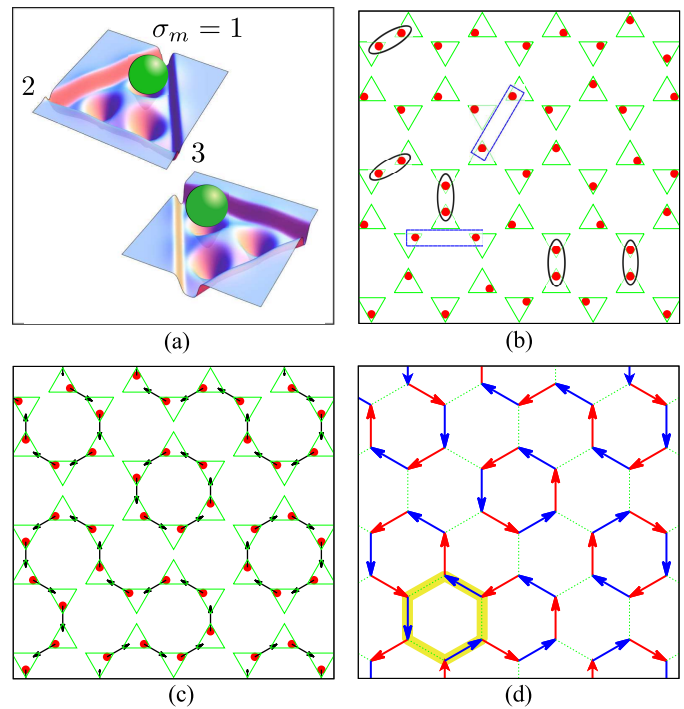}
\caption{(a) Schematic diagram of the basic unit cell with two triple well traps each containing one colloidal particle. (b) and (c) are snapshots of a small 
portion of the system. The green triangles represent the 
traps and the red dots denote the particles. (b) shows a random distribution of 
particles at high temperatures. 
$U_0$ pairs are circled and $U_1$ pairs are boxed.
(c) shows an example of a particle configuration 
that can be mapped to random fully packed loops in 
the hexagonal lattice, as illustrated in panel (d). 
The yellow contour in (d) corresponds to a flippable type-II loop.}
 \label{fig:lattice}
\end{figure}

In this letter we propose 
that a system of colloidal particles interacting with optical trap 
arrays can be used to realize fully packed loop models, and show 
that one of the resulting phases can be mapped to the three-coloring model. 
Loop models have been applied to a wide variety of
physical systems, ranging from polymer physics~\cite{duplantier,degennes} and 
turbulence~\cite{bernard06} to optics~\cite{oholleran} and magnetism~\cite{viret,nahum11,jaubert11},
and a physical realization of an idealized loop model
would be a major step in this field. 
Colloidal systems interacting with periodic optical arrays 
have been experimentally realized \cite{Brunner02,babic05,1,2,3,4} and
shown to exhibit novel types of orderings depending 
on the nature of the substrate \cite{Brunner02,1,2,11}. 
Beyond these static states, it is also possible to study a variety of dynamical
processes such as the motion of kinks and antikinks \cite{3,10}.  
Highly tailored optical trap arrays can be created where
the colloidal particles can sit in multiple positions in a single 
trapping site \cite{babic05,Babic2,4}, so that arrays 
where colloidal particles
can occupy one of three possible positions in a trap 
are well within current experimental capabilities.

{\em Model.} We consider a 2D array of $N$ triangular-shaped 
traps that each contain three potential minima, as illustrated in Fig.~\ref{fig:lattice}(a). 
These traps are similar to those created experimentally in Ref.~\cite{Brunner02}.
The traps form a honeycomb lattice with triangles 
of opposite orientations occupying the two different sublattices, 
as shown in Fig.~\ref{fig:lattice}(b) and (c). 
When each trap contains one colloidal particle, 
the system provides a natural realization of the 3-state
Potts model on the honeycomb lattice. We 
introduce a Potts variable $\sigma_i = 1$, 2, or 3 to denote the
potential well occupied by the particle in the
$i$-th trap. In order to characterize the colloidal ordering, 
we also introduce three unit vectors for each Potts state: $\hat{\mathbf e}_1 = (0, 1)$,
and $\hat{\mathbf e}_{2,\, 3} = (\pm\sqrt{3}/2, -1/2)$. The particle in the $i$-th trap is located at $\mathbf r_i = \mathbf R_i + \mathbf d_{\sigma_i}$,
where $\mathbf R_i$ is the center of the trap and the displacement 
$\mathbf d_{\sigma} = \pm \delta\, \hat{\mathbf e}_\sigma$,
with the plus (minus) sign for up (down) triangles, and $\delta$ denotes the linear size of the trap.

\begin{figure}
 \includegraphics[width =3.5in]{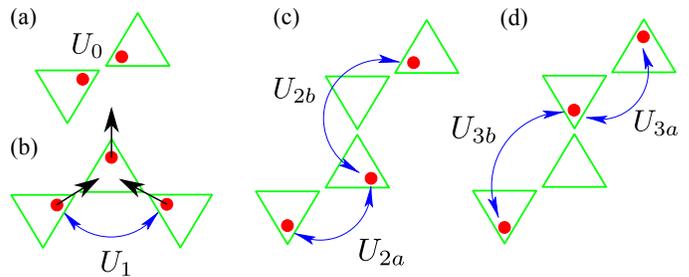}
 \caption{Various interaction terms arising from the screened-Coulomb or Yukawa potential $V(r_{ij})$ 
 between a pair of colloids in the optical traps.}
 \label{fig:interaction}
\end{figure}

The colloidal particles interact with each other via a repulsive screened Coulomb or Yukawa potential
given by $V(r_{ij}) = V_0\, q^2 \exp(-\kappa r_{ij})/r_{ij}$. Here $r_{ij}$ is the distance between two particles, 
$V_0 = {Z^{*2}}/(4\pi \epsilon\epsilon_0)$, $Z^*$ is the unit of charge, 
$\epsilon$ is the solvent dielectric constant, $q$ is the 
dimensionless colloid charge, and $1/\kappa$ is the screening length.
The Hamiltonian of the model system reads
\begin{eqnarray}
	\label{eq:H}
	\mathcal{H} = \frac{1}{2}\sum_{i,j} V\left(|\mathbf R_{ij} + \mathbf d_{\sigma_i} - \mathbf d_{\sigma_j} |\right)
\end{eqnarray}
where the summation runs over all pairs of triangular traps $i, j$.
Since the particles are always confined to one of the three potential 
wells in each trap, we can identify 
the first few neighboring interaction terms
of the colloidal potential $V(r_{ij})$ as
summarized in Fig.~\ref{fig:interaction}. 
The dominant $U_0$  interaction is between two particles 
at the closest corners of two neighboring triangles 
shown in Fig.~\ref{fig:interaction}(a).
Since these two potential wells are specified by the same Potts state in the respective traps, the $U_0$ term essentially introduces
an {\em anisotropic} antiferromagnetic interaction between the Potts variables:
\begin{eqnarray}
	\label{eq:H0}
	\mathcal{H}_{0} = U_0 \sum_{\langle ij \rangle} \delta_{\alpha_{ij}, \sigma_i} \,\delta_{\alpha_{ij}, \sigma_j},
\end{eqnarray}
where  $\langle ij \rangle$ denotes two 
nearest-neighbor traps, and $\alpha_{ij} = 1, 2, 3$ specifies the relevant Potts state of the adjacent wells of the $\langle ij \rangle$ pair.
Such $U_0$-pairs of particles [circled in Fig.~\ref{fig:lattice}(b)] 
are energetically unfavorable and will be suppressed at temperatures $T \ll U_0$.  
It is worth noting that the interactions in Eq.~(\ref{eq:H0}) are frustrated and there exist extensively degenerate Potts states 
(colloidal configurations without $U_0$-pairs) 
that minimize $\mathcal{H}_0$.

By attaching an arrow to each particle pointing 
from the center of the triangular trap to the 
corner occupied by the particle, the colloidal configuration
can be mapped to a collection of directed strings. 
Since the triangles form a honeycomb lattice, 
a similar mapping can be established
by extending the arrow onto the corresponding bond 
[see Fig.~\ref{fig:lattice}(c) and (d)]. 
As each trap contains exactly one particle, 
there is always an outgoing arrow for each vertex of the honeycomb lattice; 
however, the number of incoming arrows 
for individual vertices can be 0, 1, or 2. 
The number of vertices with no incoming arrow must equal the number of
vertices with 2 incoming arrows since these are the
sources and sinks (or end points) of the directed strings. 
The second and most relevant $U_1$ term of the interaction, 
shown in Fig.~\ref{fig:interaction}(b),
prevents the fusion of two strings by penalizing 
vertices with two incoming arrows. 
Examples of $U_1$ pairs are highlighted by square boxes in 
Fig.~\ref{fig:lattice}(b). The number of end point vertices is suppressed 
at temperatures $T \ll U_1$, where for systems with periodic 
boundary conditions (BC) it becomes energetically more favorable for
strings to form closed loops as shown in Figs.~\ref{fig:lattice}(c) 
and (d) \cite{blote94,kondev94}.  For finite lattices with open BC, 
the end points of the strings reside at the boundaries of the system.
The further-neighbor interactions $U_{2}$ and $U_{3}$ 
shown in Fig.~\ref{fig:interaction}(c,d) 
induce long-range ordering of particles at very low temperatures. 
In particular, the $U_{2b}$ term favors alignment of 
particles in two different alternating Potts states along one of 
the $C_3$ symmetry directions, 
effectively introducing a bending stiffness to the strings.

It is worth noting that each fully packed loop (FPL) configuration on the honeycomb can be further mapped to a 
three-colored configuration on the same lattice. In Baxter's 
three coloring model~\cite{baxter70},
each bond of the honeycomb lattice is assigned a color $\mathsf{R}$, $\mathsf{G}$, or $\mathsf{B}$, 
so that three different colors meet at each vertex,
and all such configurations are given equal statistical weight. The $\mathsf{R}$ and $\mathsf{B}$ colored bonds thus form a FPL configuration as
illustrated in Fig.~\ref{fig:lattice}(d), 
and the two different sequences  
$\mathsf{RBRB}\cdots$ and  $\mathsf{BRBR}\cdots$ correspond to the forward
and backward propagating loops, respectively. It is important to note that all three-colored configurations are energetically degenerate if we
retain interactions up to the $U_1$ terms only. Long-range orderings are induced by the further-distance interactions in $V(r_{ij})$.

To understand the various phases of the model system, we perform Monte Carlo simulations on the effective 3-state
Potts model described by Eq.~(\ref{eq:H}). At high temperatures, the standard single-site Metropolis updates are sufficient
to equilibrate the system;  however, 
such local updates experience a dynamical 
freezing at temperatures $T \ll U_1$ due to
the huge energy cost of updating a single-site Potts state. 
Since the effective degrees of freedom in this temperature regime
are the fully packed loops discussed above, 
we also implement two types 
of non-local updates in our Monte Carlo simulations similar
to the loop algorithm introduced for the three-coloring problem~\cite{huse92}.
In the first type of loop update, we
randomly select a loop of head-to-tail arrows, 
or a $\mathsf{RBRB}~\cdots$ loop, and flip all the arrows;
this move is accepted according to the standard Metropolis conditions with 
further-distance interactions $U_2$, $U_3$, $\cdots$
taken into account.  
The type-II loops consist of alternating bonds with and without arrows; they correspond to the
$\mathsf{GB}$ or $\mathsf{GR}$-colored loops in the 3-color scheme. 
An example type-II loop is shown in Fig.~\ref{fig:lattice}(d).

\begin{figure}
\includegraphics[width =3.2in]{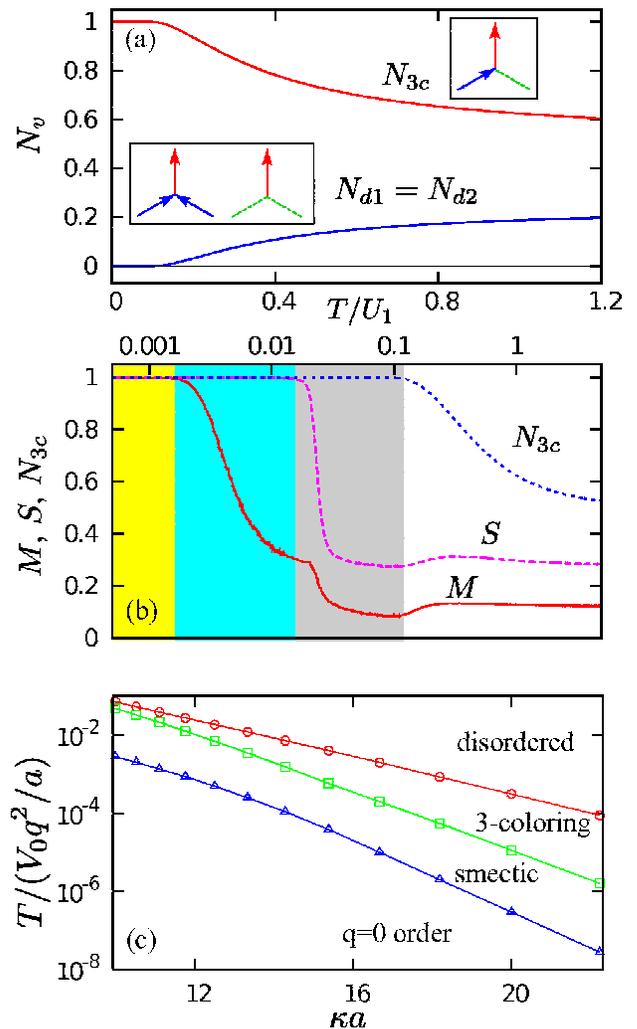}
 \caption{(a) $N_v$, the fraction of vertices of type $v$,
as a function of temperature $T/U_1$.
Upper red line: 
$N_{3c}$; 
lower blue line: 
$N_{d1}+N_{d2}$; 
dashed line: all other vertex types.
Insets: schematics of the three low-temperature vertex types
$N_{d1}$, $N_{d2}$, and $N_{3c}$.
(b) Order parameters $M$ and $S$ along with
$N_{3c}$ as a function of temperature $T/U_1$. 
The parameter $M$ characterizes a uniform long-range
ordering of particles in which 
all loops are directed 
in the same direction and parallel to each other. 
The stripe order parameter $S$
describes a partially ordered phase in which 
loops are parallel to each other but the direction of 
individual loops is disordered.
(c) Phase diagram of temperature $T$ in units of $V_0q^2/a$ 
vs $\kappa a$ showing the
regions in which the ordered, smectic, three-coloring, and disordered
states are observed.
}
 \label{fig:simulation}
\end{figure}

At very low temperatures, even the loop updates suffer freezing problems. 
Unlike the loops in dimer or spin-ice
models~\cite{barkema98,alet05}, 
which can be constructed step by step from numerous possible paths,
the loops in the three-coloring problem are predetermined by 
the colors in a given configuration. 
The so-called worm algorithm~\cite{prokofev01},
in which detailed balance is always satisfied when constructing the loop, 
cannot be applied to our case. The freezing problem arises because the acceptance rate of flipping a long loop in the standard Metropolis 
criterion becomes exceedingly small at low $T$. To overcome this problem, we employ the parallel tempering algorithm~\cite{hukushima96}
to simulate the low-temperature regime. By simultaneously simulating 150 replicas covering a temperature window $0 < T < 0.1 U_1$, 
we are able to fully equilibrate a system with periodic boundary conditions
containing $N = 2\times 6\times 12$ particles;
the results from a system with linear trap size $\delta=0.9\times (a/2\sqrt{3})$
and screening length $\kappa^{-1}=0.06 a$, where $a$ is the lattice constant
of the underlying honeycomb lattice,
are summarized in Fig.~\ref{fig:simulation}. 

In Fig.~\ref{fig:simulation}(a) we plot 
the fraction $N_v$ of honeycomb lattice
vertices of type $v$
as a function of temperature in the regime $T \ll U_0$.
Since the occurrence of $U_0$-pairs is strongly suppressed in this regime, 
there exist only three vertex types $N_{d1}$, $N_{d2}$, and $N_{3c}$,
defined according to the `coloring' of the three bonds attached
to the vertex, as
illustrated in the insets of Fig.~\ref{fig:simulation}(a). 
The three bonds meeting at the lowest-energy $N_{3c}$ vertices
always have different colors.
In the language of loops, these 3-color vertices 
have exactly one incoming and one outgoing arrow. There are 
two types, $N_{d1}$ and $N_{d2}$, 
of higher-energy defect vertices that violate 
the three-color constraints; they correspond to the 
sources and sinks of the open strings, and always satisfy
$N_{d1}=N_{d2}$.
In Fig.~\ref{fig:simulation}(a), 
as $T$ decreases the fraction of 
defect vertices $N_{d1}+N_{d2}$ gradually decreases before
vanishing for $T < T_{3c} \approx 0.1 U_1$, while
the fraction of 3-color vertices $N_{3c}$ saturates to 1
at low $T$. 
The system can thus be mapped to a three-colored or 
fully-packed loops configuration 
below the characteristic temperature $T_{3c}$.

\begin{figure}
\includegraphics[width =3.5in]{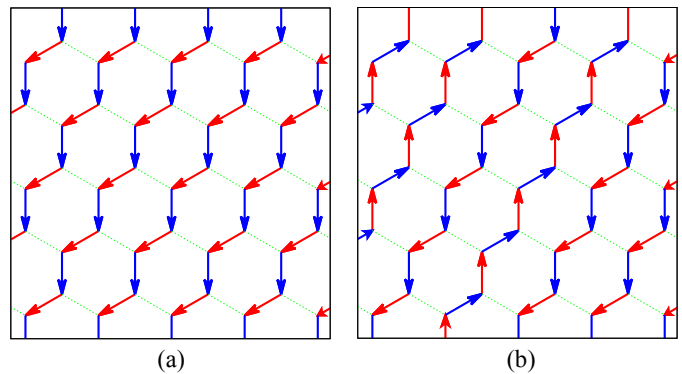}
\caption{(a) 
A long-range ordered loop state characterized by an 
antiferromagnetic Potts order parameter $M$. 
The parallel loops in this ordered state are directed in the same direction. 
(b) A partially ordered loop state exhibiting a sliding 
symmetry. The loops in this phase are parallel
to each other but the direction of individual loops remains disordered.
This state is characterized by the stripe order parameter $S$.}
\label{fig:order}
\end{figure}

As discussed previously, the further-neighbor interactions $U_2$ and $U_3$ induce long-range orderings of loops at lower
temperatures. In particular, the loops 
acquire a bending stiffness due to the $U_{2b}$ interaction. 
As a result, the loops start to
align themselves with one of the three 
principle lattice symmetry directions upon lowering the temperature.
Since the dominant Potts interaction $U_0$ is antiferromagnetic,  
we consider a N\'eel type order parameter:  $\mathbf M = (2/\sqrt{3}N) \sum_i (-1)^i \hat{\mathbf e}_{\sigma_i}$,
where $(-1)^i = +1$ for up triangles and $-1$ for down triangles. The order parameter $M = |\mathbf M|$ indeed rises to its 
maximum at $T \lesssim T_N \approx 0.003 U_1$ 
as shown in Fig.~\ref{fig:simulation}(b), indicating a ground state 
with long-range antiferro-Potts order. 
One of the perfectly ordered states 
is illustrated in Fig.~\ref{fig:order}(a);
there are a total of 6 degenerate ground states related 
to the breaking of $Z_2$ sublattice (the arrows in the loops) 
and $C_3$ rotational (the orientation of the loops) symmetries.

Interestingly, for decreasing temperature
the order parameter $M$ shows an upturn at 
$T_S \approx 0.034 U_1$, above the onset of long-range
Potts order. Examination of the snapshots from Monte Carlo simulations shows a partially ordered phase with additional sliding 
symmetries~\cite{nussinov05}. In this phase, the loops are either 
parallel or antiparallel to each other, hence breaking the $C_3$ 
lattice rotational symmetry. The directions of arrows 
in individual loops remain disordered as shown in Fig.~\ref{fig:order}(b). 
This partially ordered phase is characterized by a $Z_3$ order 
parameter indicating the overall orientation of loops and a set of Ising 
variables $\{ \tau_0, \tau_1, \cdots, \tau_L\}$  specifying 
the direction of each loop.
To characterize this stripe-like order, we first compute the 
antiferro-Potts order on a 1D chain along one of the $C_3$ axes: 
$m_{\alpha}(c) = (1/L) \mathbf v_\alpha\cdot\sum_{n \in c} (-1)^n \hat{\mathbf e}_{\sigma_n}$,
where $n$ is a site index along the chain $c$; $\alpha = 1, 2, 3$ specifying the orientation of the chains;
and $\mathbf v_{\alpha} = \hat{\mathbf e}_\beta - \hat{\mathbf e}_\gamma$, where $(\alpha\beta\gamma)$ is a cyclic permutation of $(123)$.
The vector $\mathbf v_{\alpha}$ is used to project the vector sum to the relevant Potts states along the chain.
Averaging over chains of the same orientation $\alpha$ gives a quasi-1D order parameter: $M_\alpha = (1/L) \sum_c |m_{\alpha}(c)|$, 
and finally the stripe order parameter is defined as 
their maximum $S = \max_{\alpha} M_{\alpha}$.
As shown in Fig.~\ref{fig:simulation}(b), the system enters the partially ordered stripe phase at $T \lesssim T_S$ as the
order parameter $S$ saturates to its maximum.

We summarize the sequence of thermodynamic transformations, 
illustrated in the phase diagram in Fig.~3(c), as follows. 
As the temperature is lowered, the colloidal
system first undergoes a crossover into the 3-color or random FPL phase at $T_{3c} \sim \mathcal{O}(U_1) $.
A phase transition into the partially ordered phase occurs 
at $T_S \sim \mathcal{O}(U_{2b})$ when the stripe-ordering
arises from the positive bending energy 
produced by the $U_{2b}$ interaction. 
Finally, the system undergoes another phase transition
into the long-range antiferro-Potts ordered ground state at $T_N$. 
We note that for larger system sizes, our Monte Carlo simulations combining local Metropolis, loop updates, and parallel 
tempering 
are able to reach the equilibrium 3-color phase at $T < T_{3c}$. 
However, since full equilibration to the partially ordered striped phase
as well as the fully ordered ground state requires flipping system-size 
loops, which costs too much energy for larger lattices,
our algorithm can only produce a multi-domain stripe phase.
It is worth noting that in the thermodynamic limit, the system cannot
reach the true long-range order and stays in this smectic-like phase due to 
the huge energy barrier separating different stripe states.

In summary we have proposed that 
colloidal particles interacting with a honeycomb array of optical traps that
each contain three wells
can be used to realize a fully packed loop model.
We show that this system exhibits an ordered 
ground state, a smectic-like stripe phase with a sliding symmetry, 
a random fully packed loop state, and a disordered state with broken loops.   
The random fully packed loop state can be 
mapped to Baxter's three-coloring problem, indicating that our
system could be used to create a
physical realization of this problem. We map out where these phases occur
as a function of temperature and interaction strength.   
Fully packed loops on different lattices can be similarly realized 
with optical arrays in which the number of potential wells in a trap site is 
the same as its coordination number.   
Our results should be generalizible to other systems of 
repulsively interacting particles 
in a similar array of three-well traps,
such as for vortices in BEC's interacting with optical arrays,
vortices in nanostructured type-II superconductors, 
and ions in tailored trap arrays.  

\acknowledgements
This work was carried out under the auspices of the NNSA of the U.S. DoE at
LANL under Contract No. DE-AC52-06NA25396.

\end{document}